\documentclass[runningheads,a4paper]{llncs}

\usepackage{amssymb}
\usepackage{graphicx}
\usepackage{amsfonts}
\usepackage{graphicx, epsfig,amsmath,amssymb,latexsym,graphics,float}
\usepackage{setspace,subfig}
\usepackage{citesort}
\usepackage{float}
\usepackage[ruled,vlined,linesnumbered]{algorithm2e}
\usepackage{booktabs}
\usepackage{lipsum}
\usepackage{multicol}
\usepackage{url}
\urldef{\mailsa}\path|bins@ieee.org|
\urldef{\mailsb}\path|chenkj@njupt.edu.cn|

\def\M{{\cal{M}}}

\def\v{\mathbf{v}}
\def\ind{\thicksim}
\begin{document}

\mainmatter  

\title{Maximum Likelihood Estimation based on \\
Random Subspace EDA: Application to \\
Extrasolar Planet Detection}

\titlerunning{Random Subspace EDA}

%
%
\author{Bin Liu%
\thanks{Correspondence author. E-mail: bins@ieee.org.}%
\and Ke-Jia Chen}
\authorrunning{B. Liu, et al.}

\institute{School of Computer Science, Nanjing University of Posts and Telecommunications, Nanjing,
Jiangsu, 210023, China\\
Jiangsu Key Laboratory of Big Data Security $\&$ Intelligent Processing, Nanjing, Jiangsu, 210023, China\\
}
%
%

\toctitle{Lecture Notes in Computer Science}
\tocauthor{Authors' Instructions}
\maketitle

\begin{abstract}
This paper addresses maximum likelihood (ML) estimation based model fitting in the context of extrasolar planet detection. This problem is featured by the following properties: 1) the candidate models under consideration are highly nonlinear; 2) the likelihood surface has a huge number of peaks; 3) the parameter space ranges in size from a few to dozens of dimensions. These properties make the ML search a very challenging problem, as it lacks any analytical or gradient based searching solution to explore the parameter space. A population based searching method, called estimation of distribution algorithm (EDA), is adopted to explore the model parameter space starting from a batch of random locations. EDA is featured by its ability to reveal and utilize problem structures. This property is desirable for characterizing the detections. However, it is well recognized that EDAs can not scale well to large scale problems, as it consists of iterative random sampling and model fitting procedures, which results in the well-known dilemma \emph{curse of dimensionality}. A novel mechanism to perform EDAs in interactive random subspaces spanned by correlated variables is proposed and the hope is to alleviate the \emph{curse of dimensionality} for EDAs by performing the operations of sampling and model fitting in lower dimensional subspaces. The effectiveness of the proposed algorithm is verified via both benchmark numerical studies and real data analysis.
\keywords{estimation of distribution, extrasolar planet detection, maximum likelihood estimation, nonlinear model, optimization, random subspace}
\end{abstract}

\section{Introduction}\label{sec:intro}
This paper presents an evolutionary computation based maximum likelihood (ML) estimation method for multivariate highly nonlinear time series models.
This work is motivated by a challenging signal detection task, which aims to detect extrasolar planets (exoplanets) based on observations collected by astronomical instruments such as NASA's Kepler space telescope \cite{lissauer2014advances}. The terminology ``exoplanets'' denotes planets outside our solar system. The goal of exoplanet science is to answer the scientific quest whether we are alone or whether there are other planets that might support life in the universe \cite{borucki2010kepler,lissauer2014advances,loredo2012bayesian}.
Exoplanet science has become a booming field in astrophysics since 1992 when the first detection of an exoplanet was confirmed \cite{wolszczan1992planetary}.
In this paper, we focus on the radial velocity (RV) method of exoplanet detection \cite{desort2007search,liu2014adaptive,loredo2012bayesian}. This method has become one of the most productive techniques for detecting exoplanets so far.

Signal processing (SP) plays an important part in exoplanet detection, which contributes to improve the signal-to-noise ratio
of the observations, detect signals of potential planets and so forth. Current SP techniques fall short of meeting the fundamental requirement for the future of this field. For example, the ML based periodogram method was developed to deal with correlated noise in RV time series \cite{baluev2013planetpack}, while, it is just limited to detect one signal. Many planetary systems are found to contain more than one planet, which means the RV time series should exhibit more than one signal. The Bayesian simulation techniques have been applied to explore the parameter space of a global RV model using Markov Chain Monte Carlo (MCMC) or adaptive importance sampling methods \cite{loredo2012bayesian,brewer2015fast,liu2014adaptive}, while such methods require very large computational overhead to guarantee a satisfactory performance in parameter estimation and signal detection.

The objective of this paper is to propose a computationally efficient method to address the problem of exoplanet detection based on ML fitting of complex RV models. This problem is featured by the following properties: 1) the candidate models under consideration are highly nonlinear; 2) the likelihood surface has a huge number of peaks; 3) the parameter space ranges in size from a few to dozens of dimensions. This problem lacks any analytical or gradient based searching solution to explore the parameter space. The proposed approach is based on an evolutionary computation method called estimation of distribution algorithm (EDA) \cite{zhang2004convergence,pelikan2002survey,hauschild2011introduction}. A novel mechanism is proposed to perform EDAs in a series of random subspaces spanned by correlated variables. The basic idea is that, since the dimension of each subspace will be smaller than that of the full parameter space, the number of occurrences of sample-starved model fitting will decrease and then the \emph{curse of dimensionality} is hoped to be alleviated. A benchmark numerical study and a real data analysis are used to demonstrate the effectiveness of the new algorithm.
\section{RV Models and the ML based Model Fitting}
In this section, we present the RV time series models and then introduce the ML parameter fitting problem in the context of exoplanet detection.
\subsection{RV Models}\label{sec:rv_model}
A succinct introduction of the RV models is presented here. For more details, readers are referred to \cite{liu2014adaptive}.
We use $\M_j$, $j=0,1,\ldots,J$, to denote the $j$-planet model, corresponding to the hypothesis that there is (are) $j$ planet(s) in the extrasolar system under consideration.
%
In the 0-planet model $\M_0$, the $i$th element of the RV data $v_i$ is modeled to be Gaussian distributed as follows
\begin{equation}\label{0p_Model}
v_i \mid \M_0\ind \mathcal{N}\left(C,\sigma_i^2+s^2\right),
\end{equation}
where $C$ and $\sigma^2_i + s^2$ are its mean and variance, respectively. Here $C$ denotes constant center-of-mass velocity of the star
relative to earth and $s$ denotes the square root of the ``stellar jitter'', which represents the random fluctuations in a star's luminosity or fluctuations stemming from other systematic sources, e.g., starspots. The additional variance component $\sigma^2_i$ is a calculated error of $v_i$ due to the observation procedure.

In the 1-planet model $\M_1$, $v_i$ is modeled as follows
\begin{equation}\label{Velocity_Model}
v_i \mid \M_1 \ind \mathcal{N}\left(C +\bigtriangleup
V(t_i|\phi_1),\sigma_i^2+s^2\right),
\end{equation}
where $\bigtriangleup
V(t_i|\phi)$ is the velocity shift caused by the presence of the planet. Such velocity shift is
a family of curves parameterized by a 5-dimensional vector
$\phi\triangleq(K,P,e,\omega,\mu_0)$ defined as follows
\begin{equation}\label{Velocity_1p_Model}
\bigtriangleup V(t|\phi)=K[\cos(\omega+T(t))+e\cos(\omega)],
\end{equation}
where $T(t)$ is the ``true anomaly at time $t$'' given by
\begin{equation}\label{true_anomaly}
T(t)=2\arctan\left[\tan\left(\frac{E(t)}{2}\right)\sqrt{\frac{1+e}{1-e}}\right],
\end{equation}
and $E(t)$ is called the ``eccentric anomaly at time $t$'', which is the
solution to a transcendental equation
\begin{equation}\label{transcendental_equation}
E(t)-e\sin(E(t))=\mbox{mod}\left(\frac{2\pi}{P}t+\mu_0,2\pi\right).
\end{equation}
In the above expressions, $K$ denotes the velocity semi-amplitude, $P$ the orbital period, $e$ the eccentricity
$(0\leq e \le 1)$, $\omega$ the argument of periastron $(0\le \omega
\le 2\pi)$, and $\mu_0$ the mean anomaly at time $t=0$, $(0\le \mu_0
\le 2\pi)$. The parameters $C$, $K$ and $s$ have the same unit as velocity;
the velocity semi-amplitude $K$ is usually restricted to be
non-negative to avoid identification problems; $C$ may be
positive or negative. The eccentricity parameter $e$, $0\leq e<1$, is unitless,
with $e=0$ corresponding to a circular orbit, and larger $e$ means more eccentric orbits. Periastron is the point at which the planet
is closest to the star and the argument of periastron $\omega$ measures
the angle at which we observe the elliptical orbit. The mean anomaly $\mu_0$ is
an angular distance of a planet from periastron.

More generally, a $j$-planet model ($j\geq1$) represents the expected velocity by $C
+ \bigtriangleup V(t_i|\phi_1,\ldots,\phi_j)$, in which the overall velocity shift
$\bigtriangleup V$ takes the form of the summation of velocity shifts of each individual planet, i.e.,
\begin{equation}\label{Velocity_2p_Model}
\bigtriangleup V(t_i|\phi_1,\ldots,\phi_j)=\sum_{a=1}^j
\bigtriangleup V(t_i|\phi_a).
\end{equation}
Therefore the parameter dimension of a $j$-planet model is $2+5j$. So the more planets covered by the model, the higher dimensional it is.
Moreover, the RV models $\M_1, \M_2, \ldots$ are highly nonlinear due to the velocity shift item $\bigtriangleup V$ included in these models.
\subsection{ML Parameter Estimation of RV Models}
Here we treat exoplanet detection as a problem of model selection, which means, given a set of RV observations $\v\triangleq(v_1, \ldots, v_n)$, how to select from the candidate models $\{\M_0,\M_1,\ldots,\M_j\}$ the one that fits the data best in terms of Bayesian criterion. A full Bayesian solution needs to calculate the marginal likelihood of each candidate model, which involves large scale stochastic integrations over the whole parameter space of the candidate models \cite{loredo2012bayesian,liu2014adaptive}. It is computationally expensive to solve such stochastic integrations. Here we resort to the Bayesian information criterion (BIC) to evaluate the fitness of the candidate models to the data. A Bayesian argument for adopting BIC was presented in \cite{schwarz1978estimating}. We use $\theta_i$ and ${\Theta_i}$ to denote the parameters of $\M_i$ and the corresponding value space, respectively. The BIC metric of $\M_i$ is defined as
\begin{equation}\label{eqn:BIC}
\mbox{BIC}_i=-2\cdot\ln\hat{L}_i+k_i\cdot\ln(n)
\end{equation}
where $k_i$ is the number of free parameters to be estimated for $\M_i$, $\hat{L}_i$ is the maximal likelihood function value associated with $\M_i$, i.e., $\hat{L}_i=p(\v|\hat{\theta}_i,\M_i)$, where $\hat{\theta}_i$ are the parameter values that maximize the likelihood function, namely
\begin{equation}\label{eqn:ML}
\hat{\theta}_i={\underset{\theta\in\Theta_i}{\arg\max}}\quad p(\v|\theta,\M_i).
\end{equation}
Given a finite set of models, the model with the lowest BIC value is preferred, according to the BIC criterion. Since the second item on the right-hand side of Eqn. (\ref{eqn:BIC}) is a constant given the model, calculation of $\mbox{BIC}_i$ then translates to how to solve the maximization problem defined in Eqn. (\ref{eqn:ML}).
\section{General EDA Procedure}
To address an optimization problem such as that defined in Eqn. (\ref{eqn:ML}), a general EDA procedure typically works with a population of candidate solutions defined over the full parameter space. The initial population is generated according to the uniform distribution over all admissible solutions. The fitness function gives a numerical ranking for each solution. Here the likelihood function $p(\v|\theta,\M_i)$ plays the role of a fitness function. A subset of the most promising solutions is selected by the \emph{selection} operator. A commonly used selection operator selects a certain proportion, e.g., the best 50\% of solutions. A probabilistic model is then constructed to estimate the probability distribution of the selected solutions. Given the above model, the algorithm generates new solutions by sampling the distribution defined by the model.
The new population of solutions replaces the old population and then the modeling and sampling procedure is repeated until some termination criteria are met. The main scheme for an iteration of the EDA method is summarized as follows: starting from a population of solutions $P$,
\begin{itemize}
\item Select a population of promising solutions $S$ from $P$;
\item Build a probabilistic model $M$ from $S$;
\item Sample new candidate solutions $Q$ based on $M$;
\item Replace the old population with the new population, namely set $P$ to be $Q$.
\end{itemize}
For more details about the EDA algorithm, readers are referred to \cite{hauschild2011introduction}.
As an iterative sampling and modeling procedure, the general EDA method suffers from the well-known dilemma \emph{curse of dimensionality}.
Specifically speaking, with the increase in the dimension of the solution space, the volume of the space increases so fast that the available data points used for constructing the model become sparse, making the resulting model not qualified for guiding the searching process to find better solutions.
\section{Random Subspace EDA (RS-EDA)}
\begin{figure}[]
\begin{tabular}{c}
\centerline{\includegraphics[width=3.5in,height=1.5in]{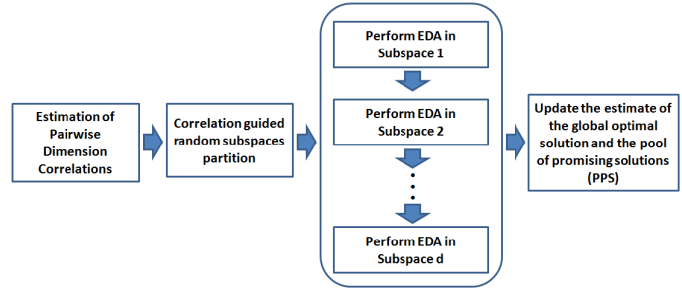}}
\end{tabular}
\caption{A conceptual scheme for one iteration of RS-EDA} \label{alg_scheme}
\end{figure}

In this section, we propose a new EDA algorithm, namely RS-EDA.
The idea is to partition the original multivariate parameter space into a series of random subspaces, and then perform EDAs in these subspaces, rather than the full parameter space, with the hope to alleviate the \emph{curse of dimensionality} for EDA type methods. Fig.\ref{alg_scheme} shows one iteration of the RS-EDA method, wherein $d$ denotes the dimension of $\theta$.
The iteration ends when the estimate of the global optimal solution keeps unchanged for a fixed number of iterations. Taking the maximization problem as an instance, we describe in what follows the details of the four operators included in Fig.\ref{alg_scheme}.
\subsection{The 1st Step: Estimation of Variable Correlations}\label{sec:cor}
This step corresponds to the leftmost box in Fig.\ref{alg_scheme}. The purpose is to provide a rough estimate on correlations among the variables (or dimensions) of $\theta$. Such estimation is made based on a pool of promising solutions (PPS), given by the previous one iteration of the RS-EDA. If it is currently in the first iteration, we initialize the PPS based on available prior knowledge on the variable correlations. If there is no such prior knowledge, we just draw a number of random samples uniformly from the solution space and then initialize the PPS via an operation of truncation selection. Specifically, we preserve the fittest 20\% of the sample for use and throw away the others.
The data in PPS is formulated by an $a\times d$ matrix, where $a$ denotes the number of samples in PPS.
The operation of correlation estimation returns the sample linear partial correlation coefficients between pairs of variables in PPS, controlling for the remaining variables in PPS. We use MATLAB's built-in function ``partialcorr'' to perform the above operation. The output of this function is a symmetric $d\times d$ matrix $C$, whose $(i,j)$-th entry is the sample linear partial correlation corresponding to the $i$th and $j$th columns of the PPS matrix.
\subsection{The 2nd Step: Random Subspace Partition}
This step corresponds to the second box in Fig.\ref{alg_scheme}.
The purpose is to partition the whole parameter space into subspaces based on the correlation matrix $C$ given by the 1st step.
This 2nd step consists of a series of procedures performed on each row of $C$.
Take the procedures corresponding to $C_i$, the $i$th row of $C$, for example.
We first sort the elements of $C_i$, namely $\{C_{i,1},\ldots,C_{i,d}\}$, according to their values from large to small. This procedure outputs $\{C_{i,j1},\ldots,C_{i,jd}\}$, where $\{j1,\ldots,jd\}$ is a rearrangement of $\{1,\ldots,d\}$ with $C_{i,jm}\geq C_{i,jn}$ as long as $m<n$. Let $S=\sum_{m=1}^dC_{i,jm}$ and then set $C_{i,jm}=C_{i,jm}/S$ for each $m$ in $\{1,\ldots,d\}$. Now we have $\sum_{m=1}^dC_{i,jm}=1$. Then we set $C'_{i,m}=\sum_{k=1}^mC_{i,jk}$ for $m=1,\ldots,d$, draw a random number $r$ from a uniform distribution between 0 and 1, find the minimum index $m$ from $\{1,\ldots,d\}$ that satisfies $C'_{i,m}\geq r$, and finally return the indexes $\{j1,\ldots,jm\}$, which we use to constitute the coordinates of the $i$th subspace.
After traversing each row of $C$, we build up a series of overlapped random subspaces.

The basic idea underlying the above operations is that the more correlation between a pair of variables, the more likely they will be incorporated into the same lower dimensional subspace. Employing the random mechanism presented above, we do not need to introduce any free parameter, such as a threshold, to cluster the variables into different subspaces.
\subsection{The 3rd Step: Performing EDAs in Subspaces}
This step corresponds to the elliptical box in Fig.\ref{alg_scheme}. Now we focus on a specific subspace and present the EDA operations endowed with it. Assume that the current estimate of the global optimal solution takes value at $\hat{\theta}=(\hat{\theta}_1,\ldots,\hat{\theta}_d)$ and the subspace under consideration is associated with variables
$(\theta_{j1},\ldots,\theta_{jm})$.
We use a Gaussian model to fit the PPS data mapped to this subspace. Specifically, we calculate the empirical mean and sample covariance of this Gaussian model based on the PPS data mapped to the dimensions $\{j1,\ldots,jm\}$. Then we draw $R\times m$ new samples from this Gaussian distribution, where $R$ is a constant prescribed beforehand. As the dimensions of all the exoplanet models of our concern are less than 100, we select $R=100$ in our experiments, which is big enough to prevent from sample-starved model fitting in the follow-up EDA operations. We assign these newly generated samples' values to the corresponding variables $\{j1,\ldots,jm\}$ of $\hat{\theta}$ and keep the other variables unchanged. Then we get a set of full dimensional samples. We calculate the fitness values of these samples, based on which we select 20\% fittest samples for use in updating the Gaussian model. During the above process, once better solutions are found, we shall update the estimate of the global optimal solution and the PPS accordingly to guarantee that the PPS maintains the fittest samples that have been found so far. Then we iterate the model fitting, sampling and selection procedure until the estimate of the global optimal solution keeps unchanged in the most recent continuous five iterations. We regard this phenomenon as an indication of algorithm convergence.
\subsection{The 4th Step: Updating the estimate of the global optimum and the PPS}
This step occupies the rightmost box in Fig.\ref{alg_scheme} for ease of presentation, while, in practice, it is totally interactive with the 3rd step presented above. Once better solutions have been found in an EDA procedure included in the 3rd step, the estimate of the global optimal solution and the PPS will be updated accordingly. In this way, the PPS always keeps a set of the fittest solutions. The EDA procedure of the next subspace will be performed based on the most recent estimate of the global optimal solution and the most recently updated PPS.
After carrying out the EDA procedures of all subspaces, the finally outputted PPS will then act as the input for the 1st step in the next iteration.
\section{Numerical Study with Benchmark Function}\label{sec:simu}
To test the potential of our idea and the ability of our algorithm to improve the scalability of EDAs to search a near-optimal
solution in large-scale problem settings, we tested it based on a benchmark function listed in
the suite of benchmark test functions released by a special session on real-parameter optimization of the 2013 Congress on Evolutionary Computation \cite{liang2013problem}. We selected the Rotated Schaffers F7 (RSF7) function, which is multimodal, nonseparable, asymmetrical, and has a huge number of the local optimum. It is defined as follows \cite{liang2013problem}
\begin{equation}
f(\theta)=\left(\frac{1}{d-1}\sum_{i=1}^{d-1}\left(\sqrt{z_i}+\sqrt{z_i}\sin^2\left(50z_i^{0.2}\right)\right)\right)^2-800
\end{equation}
where $z_i=\sqrt{y_i^2+y_{i+1}^2}$ for $i=1,\ldots,d$, $y=\Lambda_d^{10}\mbox{M}_2T_{asy}^{0.5}(\mbox{M}_1(\theta-o))$.
In the above expressions, $o$ is the shifted global optimum, $\Lambda_d^{\alpha}$ denotes a $d$ dimensional diagonal matrix, the $i$th diagonal element of which is $\alpha^{\frac{i-1}{2(d-1)}}$, $i=1,2,\ldots,d$. $T_{asy}^{\beta}(x)$ is an operator that transforms $x_i$ to be $x_i^{1+\beta\frac{i-1}{d-1}\sqrt{x_i}}$ for $x_i>0$, $i=1,2,\ldots,d$. $\mbox{M}_1$ and $\mbox{M}_2$ are orthogonal matrices whose entries are standard normally distributed. More details about this function can be found in \cite{liang2013problem}.
\begin{figure}[]
\begin{tabular}{c}
\centerline{\includegraphics[width=3.5in,height=2.2in]{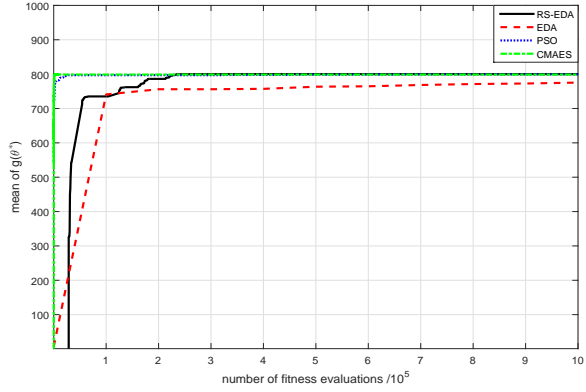}}
\end{tabular}
\caption{Algorithm comparison in a maximization task using a 5-dimensional benchmark function $g(\theta)$} \label{comp_5Dbenchmark}
\end{figure}

The RSF7 function is designed for minimization tasks. We consider maximizing the function $g(\theta)=-f(\theta)$ in order to mimic the ML estimation problem. The maximum function value of $g(\theta)$ is 800, which is the target for the algorithm to search.
The first instance we considered was a low dimensional case, in which $d$ was set at 5. We initialized the PPS of RS-EDA using $N = 20000$ random samples and we allowed a fixed budget of $10^6$ function evaluations. We compared RS-EDA with other three leading evolutionary computation methods, namely the Trelea type vectorized Particle Swarm Optimization (PSO) algorithm \cite{trelea2003particle}, the Covariance Matrix Adaptation Evolution Strategy (CMAES) method \cite{ros2008simple} and a classical EDA method termed EMNA$_{\mbox{global}}$ \cite{larranaga2002estimation}. The difference between EMNA$_{\mbox{global}}$ and our method is that the former always employs a full-dimensional multivariate Gaussian distribution model, while the latter performs EDA operations in subspaces. The population size of EMNA$_{\mbox{global}}$ is set at $N = 20000$. The population size of the PSO is set at $1000$. The CMAES method of \cite{ros2008simple} was included in our comparison because it is developed to handle high dimensional problems and it currently represents the gold-standard for comparisons in new EDA research. We used the Matlab implementation available from the authors with the diagonal option, default parameter settings and random initialization \cite{cmaesweb}. We ran all these baseline methods with a maximum number of, i.e., $10^6$, function evaluations, the same as for the RS-EDA method. We ran each method for 50 times and calculated the mean of its estimate on the maximal function value.
Fig. \ref{comp_5Dbenchmark} gives a visual summary of the results obtained in comparison. It is shown that the proposed RS-EDA method finds the global optimum with a slower convergence speed than PSO and CMAES and a faster convergence speed than EDA.

\begin{figure}[]
\begin{tabular}{c}
\centerline{\includegraphics[width=3.5in,height=2.2in]{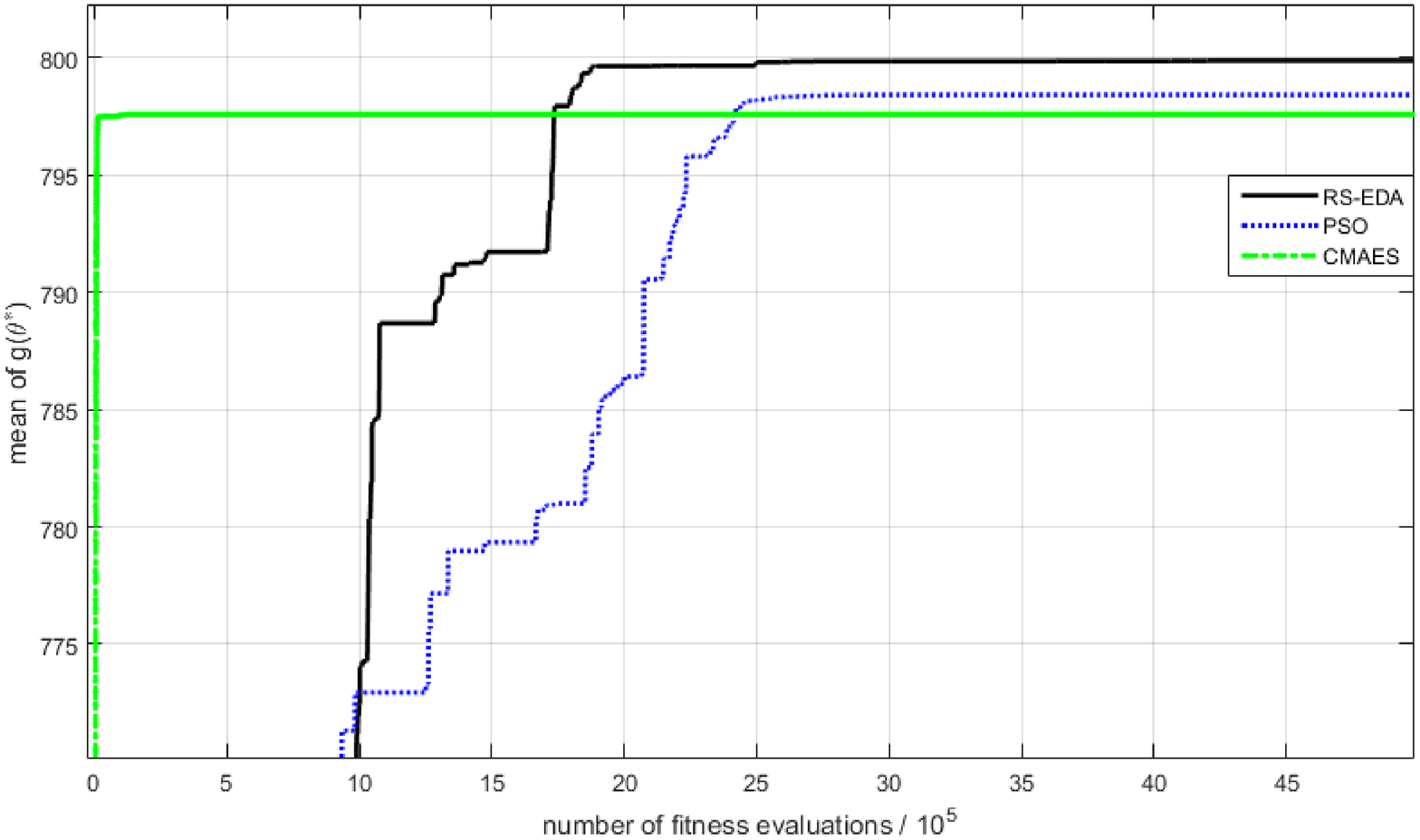}}
\end{tabular}
\caption{Algorithm comparison in a maximization task using a 20-dimensional benchmark function $g(\theta)$} \label{comp_20Dbenchmark}
\end{figure}

We then focused on a higher dimensional instance with $d$ set at 20. Compared with the first instance, the population size of each method involved increased by 10 times. The maximum function evaluations allowed for each method is $5\times 10^6$. Fig. \ref{comp_20Dbenchmark} shows the summary of the results corresponding to 50 independent runs of each method. The result of the EDA method totally diverged, so it is not included in Fig. \ref{comp_20Dbenchmark}. We see that the RS-EDA beats all the other methods.
Combining the two instances for a joint analysis, we see a potential of the proposed idea of random subspace in improving the scalability of EDAs in dealing with higher dimensional problems. However, it is worthy of further investigation.
\section{Analysis of Real Exoplanet Data }\label{sec:real_data}
We applied the proposed RS-EDA method to analyze a real data set released in \cite{tinney20062}. This data set was claimed by the astronomers to have two planets \cite{tinney20062}. An adaptive annealed importance sampling (AAIS) method was developed in \cite{liu2014adaptive}, which calculates out the probabilities of the hypothetical models based on the given RV measurements.

\begin{figure}[!htb]
\begin{tabular}{c}
\centerline{\includegraphics[width=3.5in,height=2.2in]{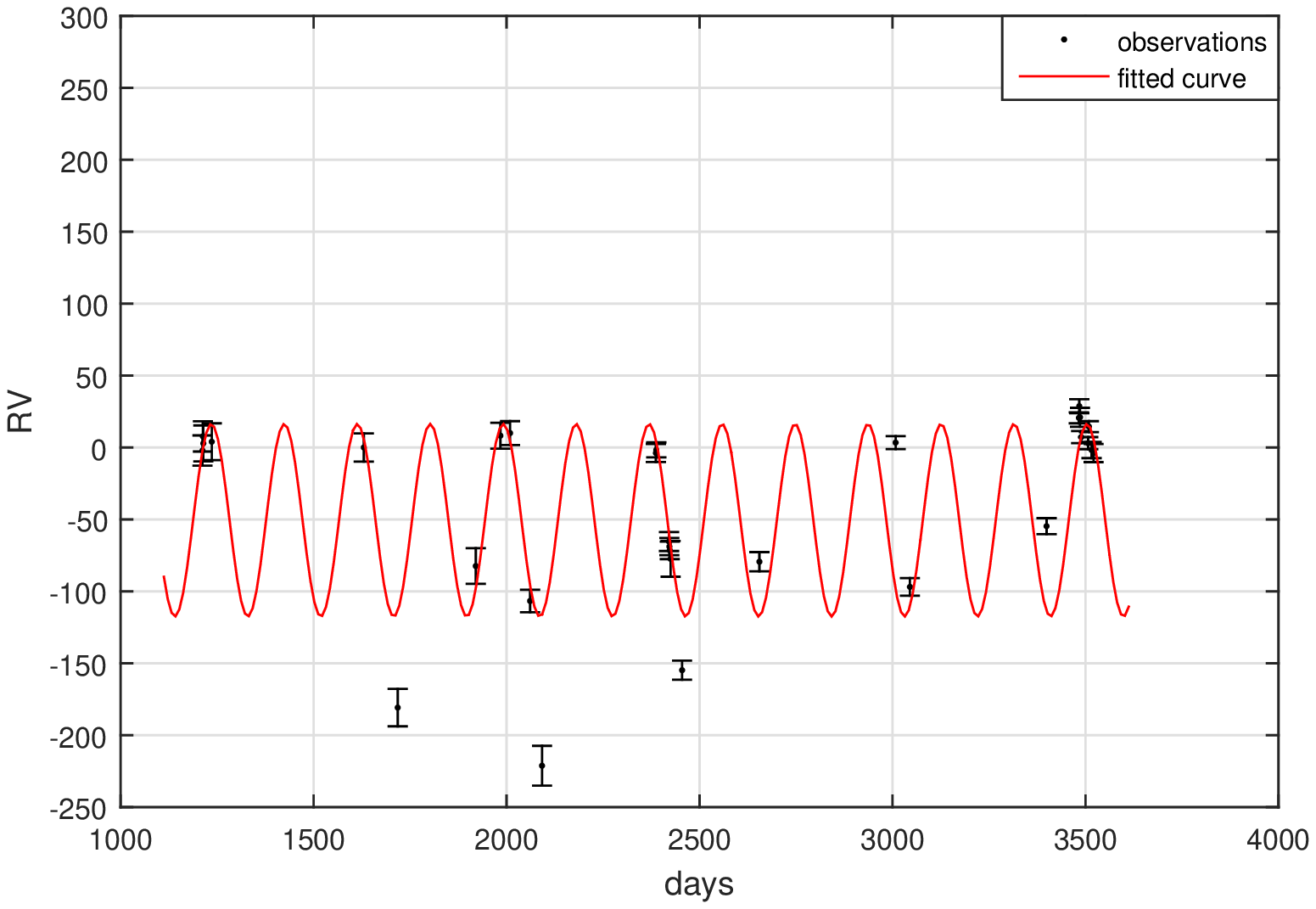}}\\
\centerline{\includegraphics[width=3.5in,height=2.2in]{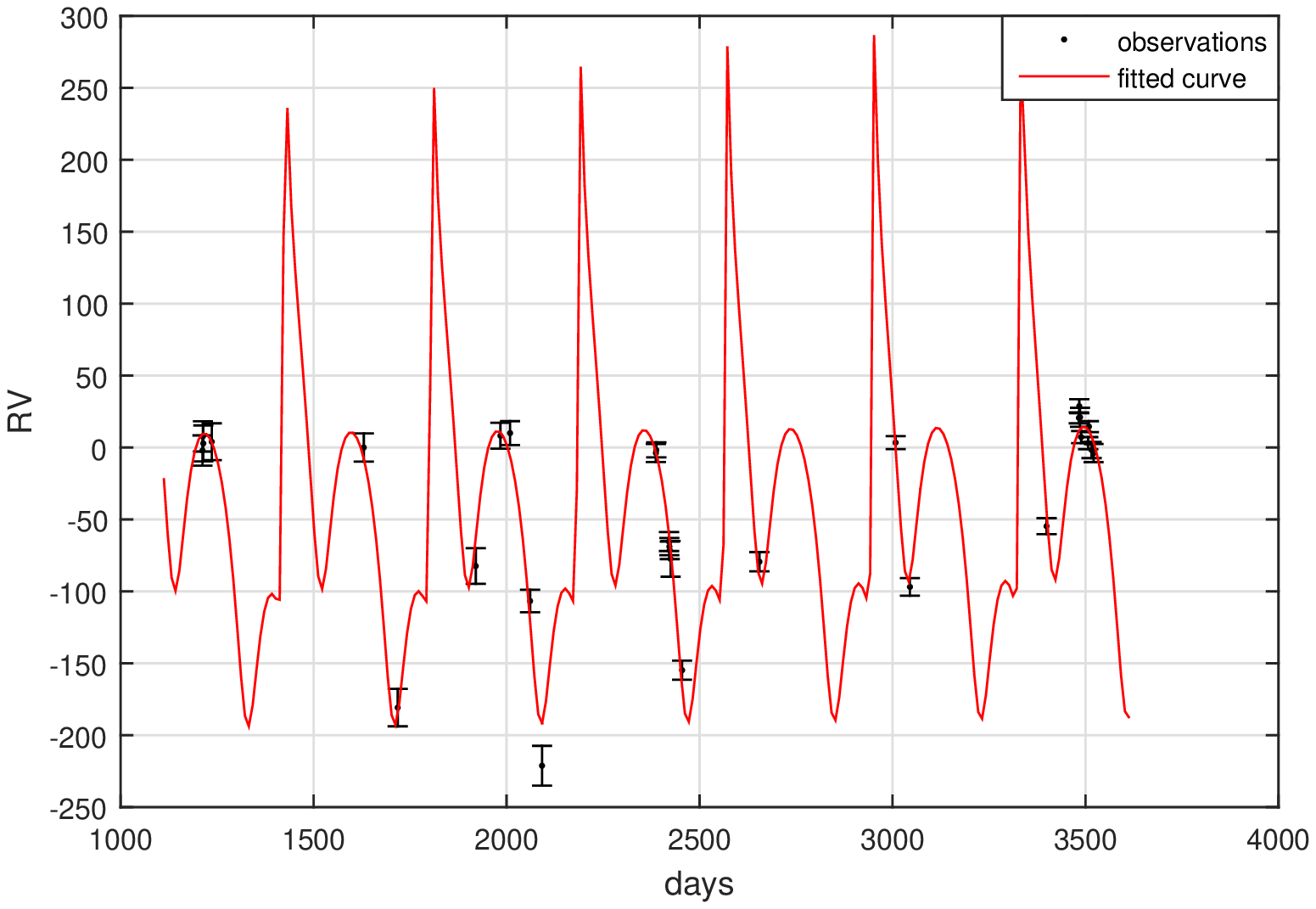}}
\end{tabular}
\caption{Measured RV data released in \cite{tinney20062} and two ML fits to the data based on $\M_1$ (the top panel) and $\M_2$ (the bottom panel), respectively. The ML solution is provided by the proposed RS-EDA method.} \label{fitting_real_data}
\end{figure}

We fit the data based on $\M_1$ and $\M_2$, respectively. For each model, we apply RS-EDA to estimate the ML model parameters.
The search space is constrained by the value ranges of the model parameters as listed in Table \ref{tab:hyper}. These ranges are provided by the astronomers based on their experiences \cite{liu2014adaptive}. Fig. \ref{fitting_real_data} shows the fitting results based on $\M_1$ and $\M_2$, respectively. The calculated logarithm of MLs and BIC metrics associated with $\M_1$ and $\M_2$ are listed in Table \ref{tab:BIC}.
Both the visual fitting result and the quantitative comparison of the BIC metrics suggest that $\M_2$ holds. This result is consistent with that reported in \cite{tinney20062} and \cite{liu2014adaptive}.
\begin{table}[!htb]
\caption{Parameter value ranges of RV models}
  \begin{center}
  \begin{tabular}{|lr|lr|} \hline
$P_{\min}$ &  $1$ day &
$P_{\max}$ &  $1,000$ years \\
$K_{\min}$ &  $ 1$ m/s &
$K_{\max}$ &  $2128$ m/s \\
$C_{min}$ &  $-2128$ m/s &
$C_{max}$ &   $2128$ m/s \\
$s_{\min}$ &  $1$ m/s &
$s_{\max}$ &  $2128$ m/s \\ \hline
  \end{tabular}
  \end{center}
\label{tab:hyper}
\end{table}

\begin{table}[!htb]
\caption{The calculated BIC metrics of $\M_1$ and $\M_2$}
  \begin{center}
  \begin{tabular}{|c|c|c|} \hline
       & $\ln(\mbox{ML})$ & BIC \\\hline
$\mbox{One Planet Model}:\M_1$ &  -148.4024  & 320.6132 \\\hline
$\mbox{Two Planet Model}:\M_2$ &  -111.5458 & 263.9060 \\
 \hline
  \end{tabular}
  \end{center}
\label{tab:BIC}
\end{table}
\section{Concluding remarks}
In this paper, we proposed an RS-EDA algorithm in the context of exoplanet detection based on ML estimation.
The most important feature of the RS-EDA method lies in the proposed operation of constructing random subspaces and then endowing the routine sampling and model fitting operations of EDAs into the lower dimensional parameter spaces, with the hope to alleviate the \emph{curse of dimensionality} for EDA type methods. The effectiveness of the proposed method was demonstrated by numerical studies and real data analysis. The results show a potential of the proposed technique in dealing with complex nonlinear models with multimodal likelihood functions and in solving high dimensional optimization tasks.
The scalability of the proposed method and new approaches for constructing random subspaces are both worthy of further investigations. 
\section*{Acknowledgement}
This work was partly supported by the National Natural Science Foundation (NSF) of China under grant No. 61571238, China Postdoctoral Science Foundation under grant Nos. 2015M580455 and 2016T90483, the Six Talents Peak Foundation of Jiangsu Province under grant No. XYDXXJS-CXTD-006 and the Scientific and Technological Support Project (Society) of Jiangsu Province under grant No. BE2016776.
\bibliographystyle{IEEEtran}
\bibliography{mybibfile}
\end{document}